\documentclass[letter, 10pt]{IEEEtran}

\usepackage{array}
\usepackage{amsmath}
\usepackage{amssymb}
\usepackage{graphicx}
\usepackage{mathtools}
\usepackage{hhline}
\usepackage{subcaption}[position=b, textformat=period]
\usepackage{makecell}
\usepackage{enumitem}
\usepackage[binary-units=true]{siunitx}
\usepackage{optidef}
\usepackage{hyperref}
\usepackage[table]{xcolor}
\usepackage{multirow}
\usepackage{booktabs}
\usepackage[ruled, linesnumbered]{algorithm2e}
\usepackage{siunitx}

\DeclareSIUnit{\belmilliwatt}{Bm}
\DeclareSIUnit{\dBm}{\deci\belmilliwatt}
\DeclareSIUnit{\dBmhz}{\dBm\per\hertz}

\newcommand\notype[1]{\unskip}
\graphicspath{{Figures/}}
\newcommand{\stateset}{\mathcal{S}}
\newcommand{\actionset}{\mathcal{A}}
\newcommand{\decisionqueue}{\mathbf{D}}
\newcommand{\packingqueue}{\mathbf{P}}

\newcommand{\cluster}{k}
\newcommand{\decisionepoch}{\tau}
\newcommand{\rstate}{S}
\newcommand{\state}{\rstate_\decisionepoch}
\newcommand{\decisionseq}{\mathbf{a}}
\newcommand{\action}{\mathbf{A}}
\newcommand{\optimalpolicy}{\pi^{*}}
\newcommand{\targetpolicy}{\pi'}
\newcommand{\reward}{R}
\newcommand{\totalclu}{K}
\newcommand{\follower}{v}
\newcommand{\packingepoch}{\widehat{\tau}}

\newcommand{\criticnet}{\hat{Q}}
\newcommand{\targetQ}{Q'}
\usepackage{mathtools}
\DeclarePairedDelimiter{\lrvert}{\vert}{\vert}
\DeclarePairedDelimiter{\floor}{\lfloor}{\rfloor}

\begin{document}
	\title{An Intelligent Transaction Migration Scheme for RAFT-based Private Blockchain in Internet of Things Applications}
	\author{Lu Hou~\IEEEmembership{Member, IEEE}, Xiaojun Xu, Kan Zheng~\IEEEmembership{Senior~Member,~IEEE,} Xianbin Wang~\IEEEmembership{Fellow,~IEEE}
	\thanks{This work was supported by the National Natural Science Foundation of China (No. 62001052).}
	\thanks{L.~Hou, X.~Xu, and K.~Zheng are with the Intelligent Computing and Communications (IC$^2$) Lab, Key Lab of Universal Wireless Communications, Ministry of Education, Beijing University of Posts \& Telecommunications, Beijing, China, 100088.}
	\thanks{X.~Wang is with the Department of Electrical and Computer Engineering, University of Western Ontario, London, ON N6A 5B9, Canada.}
	
	\thanks{Corresponding author: Kan~Zheng.}
	\thanks{Contact email: zkan@bupt.edu.cn}}

	\maketitle
	
	\begin{abstract}
	The integration of multi-access edge computing (MEC) and RAFT consensus makes it feasible to deploy blockchain on trustful base stations and gateways to provide efficient and tamper-proof edge data services for Internet of Things (IoT) applications. However, reducing the latency of storing data on blockchain remains a challenge, especially when an anomaly-triggered data flow in a certain area exceeds the block generation speed. This letter proposes an intelligent transaction migration scheme for RAFT-based private blockchain in IoT applications to migrate transactions in busy areas to idle regions intelligently. Simulation results show that the proposed scheme can apparently reduce the latency in high data flow circumstances.
	\end{abstract}
	
	
	\section{Introduction}
	Blockchain is regarded as the emerging technology that provides transparent and secure data storage for the Internet of Things (IoT) applications~\cite{blockchainiiot}. With the ability of multi-access edge computing (MEC), the gateways and base stations can participate in blockchain networks and provide tamper-resistant data services over the edge~\cite{blockchaincloud}. However, due to the limited capabilities of gateways and base stations, the ever-increasing IoT data pose a tremendous challenge in meeting the requirements of low latency~\cite{FDRL}. When an anomaly is detected in some area, the IoT system needs to increase the measurement frequency of terminals to obtain high-resolution sensing data for better decisions. It takes more time for these data to be committed into blockchain~\cite{hyperledger}.
	\par
	A lot of existed works have studied how to minimize the latency of blockchain-based IoT applications. Most of them adopt proof-of-work (PoW) consensus, and use MEC offloading to reduce the workload~\cite{offloading}. However, a large number of resources are wasted in PoW consensus to resist malicious nodes. Rovira-Sugranes \emph{et. al}~\cite{dag} optimized the latency by controlling the measurement rates of sensors in a direct-acyclic-graph (DAG)-based blockchain network. Nevertheless, DAG is not appropriate for distributed IoT applications since it makes no guarantee of consistency. As a promising consensus algorithm for distributed system, RAFT is suitable for blockchain network in IoT applications because block can be finalized by the unique leader to avoid heavy computational burden and ensure consistency. Although RAFT cannot prevent malicious node, an authentication center can be established to manage the identification and authorization of IoT devices~\cite{threat}. A few works focus on addressing issues in blockchain network with RAFT consensus. Yu \emph{et. al}~\cite{raftiiot} analyzed the latency and reliability of RAFT consensus in industrial IoT, while Xu \emph{et. al}~\cite{raftjam} studied the performance of RAFT when malicious jamming occurs in the wireless IoT environment. To the best of our knowledge, none of the literature has addressed the issue of how to reduce latency for IoT applications in RAFT-based private blockchain network. 
	\par
	Therefore, this letter proposes an intelligent transaction migration scheme by which leaders can choose to migrate transactions to other clusters optimally. The latency is defined as the sum of transactions migration latency, block generation latency and block consensus latency. The optimization problem is formulated as a Markov decision process (MDP). A deep deterministic policy gradient (DDPG) based transaction migration scheme with action refinement is proposed by which the optimal policy is generated by a neural network~\cite{DDPG, DRLauto}. The policy ensures that when the sensing data in one area increases, the incoming transactions can be migrated to other areas that are not in intensive conditions optimally. Simulations are conducted to demonstrate that the proposed scheme can reduce a lot of time for packing transactions into the blockchain.
	
	\section{System Model}
	\label{sec:model} 
\begin{figure}[t]
		\centering 
		\includegraphics[width=0.4\textwidth]{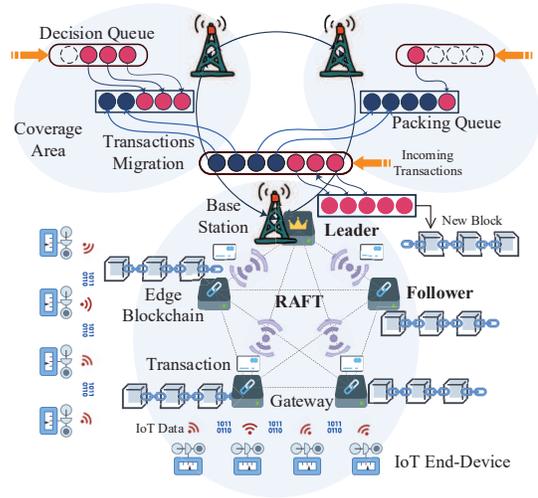}
		\caption{Illustration of RAFT-based private blockchain with transaction migration for IoT applications.}
		\label{fig:archi}
	\end{figure}

	Considering a RAFT-based blockchain system over MEC for IoT sensing application, as shown in Fig.~\ref{fig:archi}. Assume there are $\totalclu$ clusters in total, each of which maintains an independent ledger and can be indexed by $\cluster = {1, 2, \dots, \totalclu}$. Assume base stations are elected as leaders because they are more reliable than gateways in this scenario. Therefore, the term \textit{Leader $\cluster$} represents the base station in the $\cluster$-th cluster. Let $V_\cluster$ denotes the number of gateways in cluster $\cluster$. The data sensed by end-devices are sent to leaders as transactions and wait to be committed on ledgers. Let $\packingepoch$ denote the epoch when leader starts to generate new block. Assume new blocks are generated in fixed intervals and each gateway follows the leader's committing instructions and replicates the new block directly. 
	\par
	\textit{Leader $\cluster$} maintains a decision queue where transactions are waited to be processed at decision epoch $\decisionepoch$. Assume end-devices sense the environment in fixed intervals. Then, the number of arriving transactions $M_\cluster$ is constant. At one $\decisionepoch$, \textit{Leader $\cluster$} can take actions on $N$ transactions at most. The actions can be either to move transactions to the local packing queue or to migrate transactions to other clusters. Transactions in the packing queue are encapsulated in a new block at next $\packingepoch$. Let $\packingepoch = F\decisionepoch, F\in\mathbb{Z}^+$, which means block generation is triggered after $F$ decision epochs. Assume the time interval between two decision epochs is fixed and is denoted as $\Delta\tau$. Let $\decisionseq_\cluster(\decisionepoch) = \{a_{\cluster, 1}(\decisionepoch), a_{\cluster, 2}(\decisionepoch), \dots, a_{\cluster, N}(\decisionepoch)\}, \decisionepoch = 1, 2, \dots,$ denote the decision sequence of \textit{Leader $\cluster$} at $\decisionepoch$. Let $a_{\cluster, n} = 1, 2, \dots, \totalclu, 1\leq n\leq N$, where $a_{\cluster, n} = \cluster'$ means the transaction $n$ needs to be migrated from \textit{Leader $\cluster$} to \textit{Leader $\cluster'$}. Typically, $a_{\cluster, n} = \cluster$ represents that the transaction is not migrated.
	\par
	\subsection{Transactions migration latency}
	The transactions migration latency can be computed as follow,
	\begin{equation}
	t_\cluster^{\text{MIG}} = \frac{\lrvert{I_\cluster(\decisionepoch)}D}{G},
	\end{equation}
	where $I_{\cluster}(\decisionepoch) = \{n|a_{\cluster', n}(\decisionepoch) = \cluster, \cluster'\neq\cluster\}$ and $\lrvert{\cdot}$ represents the count of elements in set. $D$ is the size of a transaction and $G$ is the transmission rate of the fiber link between two leaders. The transactions are directly put into the packing queue after being migrated. No more migration is allowed to avoid high latency.
	\par
	\subsection{Block generation latency}
	At $\packingepoch$, each leader generates a new block that contains all transactions in the packing queue. A Merkle tree is established by repeated hash operations on all transactions. For $N$ transactions, $N + 2^\mathcal{N} - 1$ times of hash operations are required~\cite{Merkle}, where $\mathcal{N}\in\mathbb{Z}$ and $2^{\mathcal{N} - 1} < N \leq 2^{\mathcal{N}}$. Let $\eta_\cluster$ stands for the number of central processing unit (CPU) cycles per unit time that \textit{Leader} $\cluster$ possesses. Thus, the expected computing latency of block generation can be derived as follow,
	\begin{equation}
	t_\cluster^{\text{COM}} = \frac{(N + 2^\mathcal{N} - 1)\xi}{\eta_\cluster},
	\end{equation}
	where $\xi$ stands for the average CPU cycles needed for one hash operation.
	\subsection{Block consensus latency}
	When a block is generated, the leader needs to send the \textit{AppendEntries} message that contains the new block to all other followers to reach consensus. The transmission rate of the wireless link between \textit{Leader $\cluster$} and \textit{Follower $\follower$} is defined as follow, 
	\begin{equation}
	r_{\cluster, \follower} = \frac{B}{V_\cluster}\log(1 + \gamma_{\cluster, \follower}), 1 \leq v\leq V_k
	\end{equation}
	where $B$ is the total bandwidth, and it is equally allocated to $V_\cluster$ followers. Considering only the large-scale fading, the signal-to-noise-ratio (SNR) of the wireless link from sender $\cluster$ to receiver $\follower$ can be computed as follow,
	\begin{equation}
	\gamma_{\cluster, \follower} = \frac{Pd_{\cluster, \follower}^{-\beta}\Psi}{BN_0},
	\end{equation}
	where $P$ is the total transmission power that is equally divided for $V_\cluster$ followers. $N_0$ stands for the average power spectral density of white noise, and $d_{\cluster, \follower}$ is the distance between \textit{Leader $\cluster$} and \textit{Follower $\follower$}. $\beta$ stands for the path loss ratio, and the $\Psi\sim\mathcal{LN}(0, \sigma^2)$ denotes the shadow fading which follows log-normal distribution. 
	\par
	To reach consensus on the new block, the leader has to receive confirmation messages from at least $\zeta_\cluster + 1$ followers (including the leader itself), where $\zeta_\cluster = \floor{V_\cluster / 2}$. Therefore, the consensus latency is determined by the transmission latency caused by the leader sending \textit{AppendEntries} message to and receiving confirmation from the $\zeta_\cluster$-th follower, i.e.,
	\begin{equation}
	t_\cluster^{\text{CON}} = \frac{C}{r_{\cluster, \zeta_\cluster}} + \frac{E}{r_{\zeta_\cluster, \cluster}},
	\end{equation}
	where $C$ denotes the size of an \textit{AppendEntries} message and $E$ represents the size of the confirmation message.
	\par
	\subsection{Objective}
	Our objective is to minimize the long-term latency of all leaders by an optimal migration policy $\optimalpolicy$. Giving consideration to fairness, the long-term latency is summed by the maximal latency of all clusters. Therefore, the objective problem is defined as follow,
	\begin{mini}
		{\pi}
		{\sum_\decisionepoch \{\max_\cluster(t_\cluster^{\text{COM}} + t_\cluster^{\text{CON}} + t_\cluster^{\text{MIG}})\},}
		{\label{optimizationproblem}}
		{}
	\end{mini}
	
	\section{DDPG based transaction migration scheme}
	\label{sec:ddpg}
	The optimization problem eq.~\eqref{optimizationproblem} is formulated as an MDP model which consists of state set $\stateset$, action set $\actionset$, reward function $R: \stateset\times\actionset\rightarrow\mathbb{R}$, and transition probability $T: \stateset\times\actionset\times\stateset\rightarrow[0, 1]$. A reinforcement learning method is adopted to find the optimal policy $\pi^*$ by a training procedure. Besides, an action refinement is applied to choose the best discrete action from the original output of a neural network.
	\subsection{MDP model formulation}
	\subsubsection{\textbf{State set}}
	System state $\state$ is defined as $\state = \{\decisionqueue_\decisionepoch, \packingqueue_\decisionepoch, \Gamma_\decisionepoch, \iota\}$, where $\decisionqueue_\decisionepoch = \{D_{1, \decisionepoch}, D_{2, \decisionepoch}, \dots, D_{\totalclu, \decisionepoch}\}$ represents the number of transactions in decision queue of each leader, and $\packingqueue_\decisionepoch = \{P_{1, \decisionepoch}, P_{2, \decisionepoch}, \dots, P_{\totalclu, \decisionepoch}\}$ stands for the status of packing queue, where $P_{\cluster, \decisionepoch}$ includes transactions that come from local decision queue at previous epoch and that are migrated from other clusters, i.e., $P_{\cluster, \decisionepoch} = N - \bigr|\{a_{k, n}|a_{k, n}\neq k, n\in[1, N]\}\bigr| + \lrvert{I_{\cluster}(\decisionepoch - 1)}$. $\Gamma_\decisionepoch = \{\gamma_{1, 1, \decisionepoch}, \dots, \gamma_{\cluster, V_\cluster, \decisionepoch}, \dots, \gamma_{\totalclu, V_\totalclu, \decisionepoch}\}$ denotes the SNRs of all wireless links. An indicator $\iota = \decisionepoch \mod F$ is used to indicate the number of decision epochs after a packing epoch. If $\iota = F$, a new block is generated. All possible states form the state set $\mathcal{S}$.
	
	\subsubsection{\textbf{Action set}}
	An action $\action_{K\times N}(\decisionepoch)$ is given as a matrix, where each row stands for a decision sequence by the leader, i.e., $\action(\decisionepoch) = \{\decisionseq_\cluster(\decisionepoch)|\cluster\in[1, K]\}$. All possible actions form up an action set $\mathcal{A}$. 
	\par
	Without extra explanations, the symbol $\decisionepoch$ is ignored for simplicity, i.e.,$\state\coloneqq S$, $\decisionqueue_\decisionepoch\coloneqq\decisionqueue$, $\packingqueue_\decisionepoch\coloneqq\packingqueue$, $\Gamma_{\decisionepoch}\coloneqq\Gamma$ and $\action_{K\times N}(\decisionepoch)\coloneqq\action$.
	\par
	\subsubsection{\textbf{State transition function}}
	With an action $\action$ at a given state $\rstate$, the following state can be determined via the state transition function. According to different value of indicator $\iota$ in $\rstate$, the transition function is defined as follows,
	\begin{itemize}
		\item \textbf{When $\iota\neq F$}, transition function is given as follow,
	\begin{align}
		\rstate_{\tau + 1} &= f(\rstate, \action) = \{D_1 + M_1 - N, \dots, \\\nonumber
		&D_\totalclu + M_\totalclu - N, P_{1, \decisionepoch + 1}, \dots, P_{\totalclu, \decisionepoch + 1}, \iota + 1\},
	\end{align}
	
	\item \textbf{When $\iota = F$}, leaders need to collect all transactions in the packing queue to generate new blocks. Therefore, the transition function is given as follow,
	\begin{align}
	&\rstate_{\tau + 1} = f(\rstate, \action) = \\\nonumber
	&\{D_1 + M_1 - N, \dots, D_\totalclu + M_\totalclu - N, 0, \dots, 0, \iota = 0\},
	\end{align}
	\end{itemize}
	\subsubsection{\textbf{Reward function}}
	The instant reward is defined as the maximal latency of all leaders at $\decisionepoch$. To keep consistency with the DDPG algorithm, the value of the reward function is set to be negative to make the larger reward stand for the lower latency. The reward function is given as follow,
	\begin{equation}
		R(\decisionepoch) = \begin{cases}
			0, &\iota \neq F\\
			-\max_{\cluster\in\totalclu}\{t_\cluster^{\text{COM}} + t_\cluster^{\text{CON}} + t_\cluster^{\text{MIG}}\}, &\iota = F
		\end{cases}
	\end{equation}
	When $\iota\neq F$, no block is generated. Therefore, the reward is set to be zero. Only at $\packingepoch$ need our model to calculate latency.
	\begin{table}[t]
	\centering
	\caption{Information of neural network structure}
	\label{tab:nn}
	\begin{tabular}{c|c|c|c|c}
	\toprule
	\makecell[c]{\textbf{Network}\\\textbf{Type}} & \textbf{Layer Type} & \makecell[c]{\textbf{Number} \\\textbf{of Cells}} & \textbf{Activation} & \textbf{Loss} \\
	\midrule
	\multirow{5}{*}{} & Input Layer & \makecell[c]{$2\totalclu + 1$\\$+ \sum_k V_k$ }& - &\multirow{5}{*}{} \\
	\cmidrule{2-4}
		& Hidden Layer 1 & 8 & ReLU &\\
	\cmidrule{2-4}
	\textbf{Actor}	& Hidden Layer 2 & 16 & ReLU & $Q$\\
	\cmidrule{2-4}
		& Hidden Layer 3 & 8 & ReLU &\\
	\cmidrule{2-4}
		& Output Layer & $K*N$ & $K*\text{Sigmoid}$ & \\
	\midrule
	\midrule
	\multirow{5}{*}{} & Input Layer & \makecell[c]{$2\totalclu + 1$\\$+\sum_k V_k$\\$+K*N$ }& - & \multirow{5}{*}{} \\
	\cmidrule{2-4}
		& Hidden Layer 1 & 8 & ReLU & Mean\\
	\cmidrule{2-4}
	\textbf{Critic}	& Hidden Layer 2 & 16 & ReLU &Squared\\
	\cmidrule{2-4}
		& Hidden Layer 3 & 8 & ReLU &Error\\
	\cmidrule{2-4}
		& Output Layer & 1 & - &\\
	\bottomrule
	\end{tabular}
	\end{table}
	\subsection{Transaction migration scheme}
	Due to the large size of action space, the DDPG is adopted to make approximations on both action-value function $Q$ and policy $\pi$~\cite{DDPG}. The DDPG model includes an actor network and a critic network, each of which has a corresponding target network. Since the DDPG generates actions in continuous space, action refinement is adopted in our algorithm to choose the proper discrete action from the original continuous action, according to~\cite{actionrefinement}.
	\subsubsection{\textbf{Actor and critic networks}}
	Neural networks are used in DDPG to approximate both $Q$ and $\pi$. Let $\hat{\pi}(\rstate|\phi)$ denotes actor network with parameter $\phi$ and $\hat{Q}(\rstate, \action|\theta)$ denotes the critic network with parameter $\theta$. All layers in both networks are fully connected. The structures of target networks are equivalent to $\hat{\pi}$ and $\hat{Q}$, while the parameters are updated gradually via exponential moving average (EMA). Let $\phi'$ and $\theta'$ denote the parameters of target actor network and target critic network. Besides, let $\kappa$ denotes the rate of the EMA update on parameters. Then, the update function of target networks can be given as follow,
	\begin{equation}
	\label{eq:targetupdate}
		\phi' = \kappa\phi' + (1 - \kappa)\phi,	\theta' = \kappa\theta' + (1 - \kappa)\theta.
	\end{equation}
	For simplicity, define short notations for actor and critic networks as $\pi\coloneqq\hat{\pi}(\rstate|\phi)$, $Q\coloneqq\hat{Q}(\rstate, \action|\theta)$, and for target networks as $\targetpolicy\coloneqq\hat{\pi}(\rstate'|\phi')$, $Q'\coloneqq\hat{Q}(\rstate', \action'|\theta')$.
	\par
	The loss $J(\phi)$ of the actor network is defined as the estimated value of $Q$ that is computed by critic network, while the loss of the critic network is defined by the mean squared error of $Q$ and the target value $y$ which is given as follow,
	\begin{equation}
	\label{eq:y}
		y = \reward + \lambda Q',
	\end{equation}
	where $\lambda$ is the discount factor. Then, the loss of critic network can be computed as follow,
	\begin{equation}
		\mathcal{L}(\theta) = \mathbb{E}\{(Q - y)^2\}
	\end{equation}
	The main objective of DDPG is to find the optimal set of parameters $\theta^{*}$ to maximize $Q$ which is calculated by the long-term sum of rewards, i.e., eq.~\eqref{optimizationproblem}. 
	
	\par
	\begin{algorithm}[!t]
		\KwIn{Initialize parameters $\phi, \theta\sim\mathcal{G}$, where $\mathcal{G}$ is a Gaussian distribution. Initialize a starting state $\rstate =\rstate_0 = \{0, 0, \dots, 0\}$. Set experience deque $\mathcal{E} = \varnothing$, an index $\chi$ that indicate the epoch of training start and the total iteration times $\tau_{\max}$.}
		\KwOut{$\optimalpolicy$}
		\BlankLine
		\While{$\tau\neq \tau_{\max}$} {
			Sample a continuous action with random exploration, i.e., $\action = \pi(\rstate|\phi) + g$, where $g\sim\mathcal{G}$\;
			Refine actions according to eq.~\eqref{eq:refine}\;
			Each leader execute action $\action$ and observe the instant reward $\reward$ \;
			Environment transit to state $\rstate_{\tau + 1} = f(\rstate, \action)$ \;
			Push the observation sequence $(\rstate, \action, \reward, \rstate_{\tau + 1})$ into the experience deque $\mathcal{E}$ \;
			Set $\tau \leftarrow \tau + 1$ and $\rstate \leftarrow \rstate_{\tau + 1}$\;
			\If{$\tau > \chi$}{
				Randomly sample a mini-batch $e\sim\mathcal{E}$ \;
				Obtain target action batch by the target actor network via $\action' = \targetpolicy(\rstate)$, where $\rstate\sim e$ \;
				Acquire the target action value batch by the target critic networks via $\targetQ(\rstate, \action')$ \;
				Compute the value $y$ according to eq.~\eqref{eq:y}\;
				Calculate the gradient of the critic network as $\nabla\mathcal{L}$\;
				Update parameters $\theta$ of the critic network towards the direction of gradient descent, i.e., $\theta \leftarrow \theta - \alpha_1\nabla\mathcal{L}$, where $\alpha_1$ is the learning rate of critic network\;
				Get raw action batch by the actor network via $\action = \pi(\rstate)$, where $\rstate\sim e$\;
				Calculate the gradient of the actor network as follow,
				\begin{equation}
					\nabla J = \frac{\partial Q}{\partial\phi} = \frac{\partial Q}{\partial\action}\frac{\partial\pi}{\partial\phi}
				\end{equation}
				Update parameters $\phi$ towards the direction of gradient ascent, i.e., $\phi \leftarrow \phi + \alpha_2\nabla J$, where $\alpha_2$ is the learning rate of the actor network\;
				Update the parameters of the target networks according to eq. (\ref{eq:targetupdate})\;
			}
		}
		\caption{DDPG-based transaction migration training algorithm with action refinement}
		\label{algo:migrate}
	\end{algorithm}
	
	The detailed structure of both the actor network and the critic network is given in Tab.~\ref{tab:nn}. 
 
 		\begin{table}[!t]
	\centering
	\caption{Parameter configuration}
	\label{tab:parameter}
	\begin{tabular}{c|c|c|c}
		\toprule
		\textbf{Parameter} & \textbf{Value} & \textbf{Parameter} & \textbf{Value} \\
		\midrule
		$K$ & $4$ & $V$ & $4$\\
		\midrule
		$B$ & \SI{20}{\mega\hertz} & $N_0$ & \SI{150}{\dBm}\si{/\hertz} \\
		\midrule
		$P$ & \SI{24}{\dBm} & $\beta$ & $3.4$ \\
		\midrule
		$\xi$ & $300$ & $\eta$ & \SI{1}{\giga\hertz} \\
		\bottomrule
	\end{tabular}
	\end{table}
	
	\subsubsection{\textbf{Action refinement}}
	The output of the actor network is the continuous features of actions, which cannot be directly executed by leaders. By only using the most approaching discrete feature of action may not get the maximal $Q$ value~\cite{actionrefinement}. Therefore, an action refinement, which can choose the discrete action to reach the maximal action-value function, is given as follows,
	\begin{equation}
	\label{eq:refine}
		\action^{*} = \arg\max_{\hat{\action}\in\mathbb{A}}\criticnet(\rstate, \hat{\action}),
	\end{equation}
	where $\mathbb{A} = \{\tilde{\action}\bigr| \lVert\tilde{\action} - \pi\rVert_2\leq\delta\}$. By using the proper value of $\delta$, the balance between precision and efficiency can be reached.
	\par
	\subsubsection{\textbf{Training procedure}}
	The training process of the transaction migration algorithm is given in Algorithm~\ref{algo:migrate}. After enough training, the actor network can generate optimal transactions migration policies for each leader of the cluster.
	
	\section{Performance evaluation and analysis}
	\label{sec:simulation}
	\subsection{Simulation setup}
	A simulation is conducted to simulate the situation where the sensing frequency rises in some areas when an anomaly occurs, and descends back to normal after the anomaly disappears, in an IoT application. There are $4$ areas in total and the anomaly occurs in \textit{Area 2}. The number of incoming transactions in other areas stay to be $6$, while $M_2$ varies from $4$ to $9$ and decays back to $4$. Key parameters of our simulations are listed in Tab.~\ref{tab:parameter}. Each leader needs to receive $\zeta_k = 2$ responses from gateways to reach consensus by RAFT. Besides, the learning rates of both the actor network and the critic network are both set to be $10^{-5}$.
	
	\par

	\subsection{Numerical results and analysis}
	\begin{figure}[!t]
		\centering 
		\includegraphics[width=0.4\textwidth]{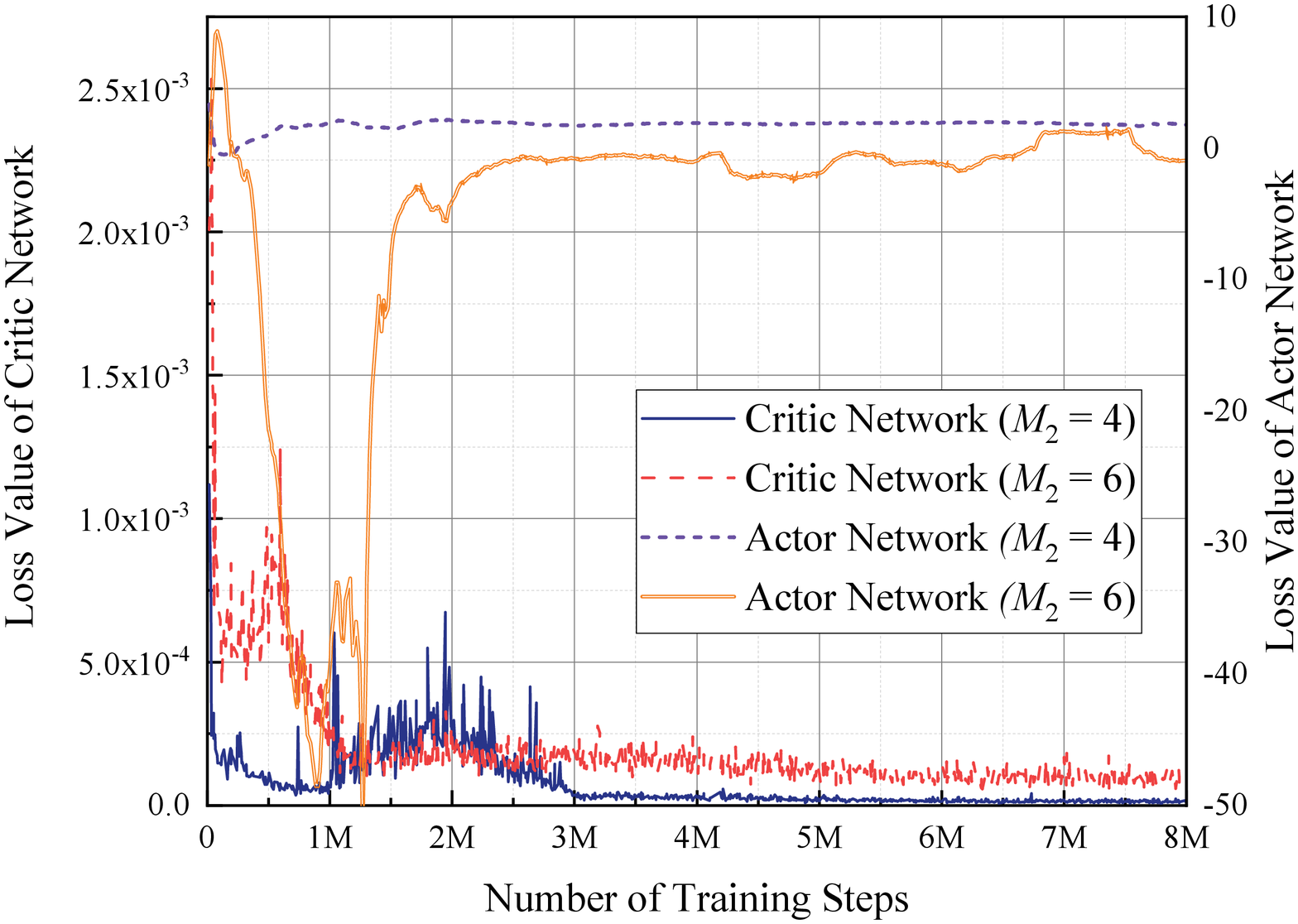}
		\caption{Illustration of convergence of actor and critic networks under different $M_2$.}
		\label{fig:conv}
	\end{figure}
	Each simulation runs for $8\times 10^{6}$ epochs. The actor loss $J(\phi)$ rises as the iteration time increases, while the critic loss $\mathcal{L}(\theta)$ is decreasing, as shown in Fig.~\ref{fig:conv}. This is because actor loss $J(\phi)$ stands for the average action value of each batch. Thus, the higher $J(\phi)$ becomes, the higher rewards can be got. On the other hand, the critic loss $\mathcal{L}(\theta)$ represents the difference between the estimated action-value function and the target action-value function. Therefore, $\mathcal{L}(\theta)$ needs to decrease during the training process.
	\par
	\begin{figure}[!t]
		\centering 
		\includegraphics[width=0.4\textwidth]{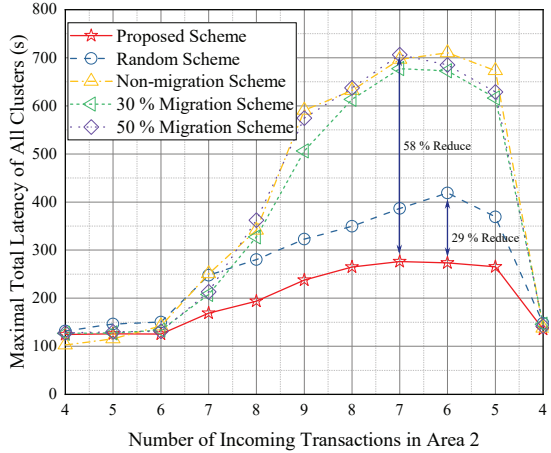}
		\caption{Illustration of latencies of all schemes under different $M_2$.}
		\label{fig:rewards}
	\end{figure}
	The proposed scheme is compared to four baselines, i.e., random scheme and three static schemes include non-migration, 30\% migration and 50\% migration schemes. As the name implies, each baseline scheme defines how many transactions are migrated. The random scheme provides the average performance for all the possible schemes since each action of this scheme can be selected with equal probability after training enough times. As shown in Fig.~\ref{fig:rewards}, the three static baselines perform bad similarly. It proves that static schemes cannot help improve the performance, even if they can migrate transactions from busy areas. Latency that is reached by the proposed scheme stays the lowest, because the proposed scheme can act intelligently with regard to the changing states. The proposed scheme balances the transactions among clusters optimally to reduce the latency by at most 58\% and 29\%, compared to the static schemes and random scheme.
	\par
\begin{figure}[!t]
		\centering 
		\includegraphics[width=0.4\textwidth]{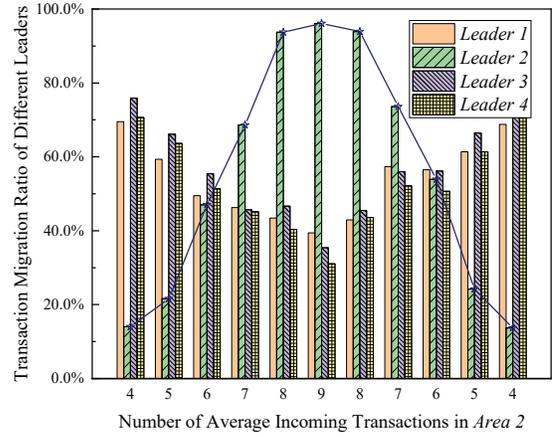}
		\caption{Illustration of migration ratios of all leaders under different $M_2$.}
		\label{fig:action_ratio}
	\end{figure}

	As shown in Fig.~\ref{fig:action_ratio}, migration ratio of \textit{Leader $2$} is the lowest at initial stage, because the workload is smaller than other three leaders. Therefore, the proposed scheme tends to migrate transactions from other leaders to reduce the maximal latency. When an anomaly occurs, the number of transactions rises in \textit{Area 2}, causing \textit{Leader $2$} to migrate transactions to other leaders as many as possible. Meanwhile, migration ratios of other leaders fall down to a low level. This is because our objective is to minimize the latency of the whole system rather than a particular cluster. Therefore, as the number of arriving transactions increases, the proposed algorithm tends to migrate transactions from the leader with high burden to the leaders that are idle. After the anomaly disappears, the migration ratios start to get back to normal. 
	
	\par
	
	\section{Conclusion}
	\label{sec:conclusion}
	In this letter, a reinforcement-learning-based scheme that can migrate transactions intelligently between clusters has been proposed for RAFT-based private blockchain in IoT application. The scheme can minimize the latency of storing IoT data on chain by balancing the workload among all clusters. DDPG algorithm with an action refinement is proposed to obtain optimal transaction migration policy on consideration of dynamic status of the environment. Simulation results show that the proposed scheme can reduce at most $58\%$ latency when the system undergoes an anomaly in some sensing area, compared to baselines.
\bibliographystyle{IEEEtran}
\bibliography{ref}

\end{document}